\documentclass[journal]{journal}
\usepackage{subfig}
\usepackage{accents}
\usepackage{amsfonts}
\usepackage{booktabs}
\usepackage{siunitx}

\usepackage{siunitx}

\usepackage{nomencl}
\makenomenclature
\newcommand{\nomunit}[1]{%
\renewcommand{\nomentryend}{\hspace*{\fill}#1}}

\usepackage{cite}

\ifCLASSINFOpdf
  \usepackage[pdftex]{graphicx}
\else
  \usepackage[dvips]{graphicx}

\fi
\usepackage{authblk}

\author[1,2,3]{Justin~Reverdi}
\author[1]{Sixin~Zhang}
\author[2]{Saïd~Aoues}
\author[3]{Fabrice~Gamboa}
\author[1]{Serge~Gratton}
\author[1]{Thomas~Pellegrini}
\affil[1]{Institut de Recherche en Informatique de Toulouse (IRIT)}
\affil[2]{LIEBHERR Aerospace Toulouse}
\affil[3]{Instiut de Mathématiques de Toulouse (IMT)}

\begin{document}
\bstctlcite{IEEEexample:BSTcontrol}

%
\title{CNN-based Compressor Mass Flow Estimator in Industrial Aircraft Vapor Cycle System}
\markboth{Journal of \LaTeX\ Class Files,~Vol.~6, No.~1, January~2007}%
{Shell \MakeLowercase{\textit{et al.}}: Bare Demo of IEEEtran.cls for Journals}
%



\newcommand{\blue}[1]{\textcolor{blue}{#1}}
\newcommand{\red}[1]{\textcolor{red}{#1}}
\newcommand{\cyan}[1]{\textcolor{cyan}{#1}}

\maketitle
\thispagestyle{empty}

\begin{abstract}
In Vapor Cycle Systems, the mass flow sensor plays a key
role for different monitoring and control purposes. However, physical
sensors can be inaccurate, heavy, cumbersome, expensive or highly sensitive to vibrations, which is especially problematic when embedded into
an aircraft. The conception of a virtual sensor, based on other standard sensors, is a
good alternative. 
This paper has two main objectives. Firstly, a data-driven model using a Convolutional Neural Network is proposed to estimate the mass flow of the compressor. 
We show that it significantly outperforms the standard Polynomial Regression model (thermodynamic maps),
in terms of the standard MSE metric and Engineer Performance metrics. 
Secondly, a semi-automatic segmentation method is proposed to compute the Engineer Performance metrics for real datasets,
as the standard MSE metric may pose risks in analyzing the dynamic behavior of Vapor Cycle Systems.

\end{abstract}


%
\IEEEpeerreviewmaketitle

\makenomenclature

\nomenclature[A]{\(\dot{m}\)}{Compressor Mass Flow
\nomunit{\unit{\kg\per\s}}}

\nomenclature[A]{\(\omega\)}{Compressor Motor Speed
\nomunit{\unit{\radian\per\second}}}

\nomenclature[A]{\(C\)}{Compressor Motor Torque
\nomunit{\unit{\N\m}}}

\nomenclature[A]{\(P_\textit{in}\)}{Compressor Inlet Pressure
\nomunit{\unit{\Pa}}}

\nomenclature[A]{\(P_\textit{out}\)}{Compressor Outlet Pressure
\nomunit{\unit{\Pa}}}

\nomenclature[A]{\(T_\textit{in}\)}{Compressor Inlet Temperature
\nomunit{\unit{\K}}}

\nomenclature[A]{\(T_\textit{out}\)}{Compressor Outlet Temperature
\nomunit{\unit{\K}}}

\printnomenclature

\section{Introduction}
%
%
%
%
 \IEEEPARstart{I}{n} physical systems, it is crucial to measure certain quantities. Traditionally, physical sensors have been used for this purpose, but they can be expensive, cumbersome, and require maintenance. Sometimes there are no physical sensors for the quantity of interest. As an alternative, virtual sensors have emerged as a cost-effective and reliable method for measuring physical data based on other physical sensors \cite{article}. 
 In complex physical systems, it is sometimes hard 
 to derive physically meaningful equations to construct a virtual sensor. 
 Therefore, using machine learning models to design a virtual sensor has become a popular 
 research direction in recent years \cite{article, reg2, cnn,8079133,9724134,DBLP:journals/corr/abs-1901-10738,9127505}. 
 In this paper, we propose and evaluate a virtual sensor based on Machine Learning (ML) for estimating an important 
 physical quantity (Compressor Mass Flow) in the Vapor Cycle System (VCS) of LIEBHERR in airplanes. 
\noindent
\\

\begin{figure}[h]
\centering
\includegraphics[width=3.2in]{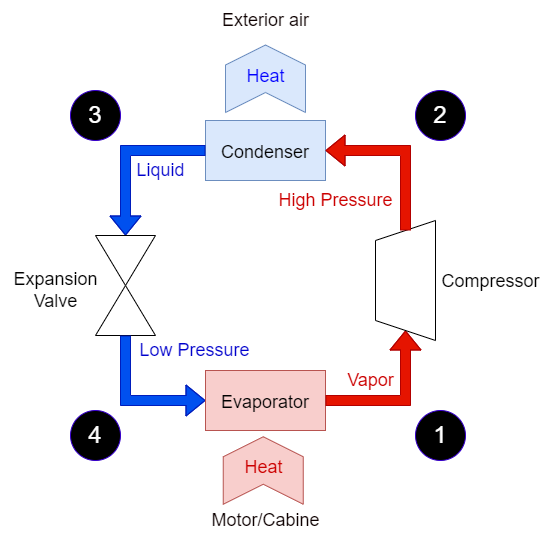}
\caption{Vapor Cycle System (VCS)}
\label{fig_sim1}
\end{figure}

\begin{figure}[h]
\centering
\includegraphics[width=3.4in]{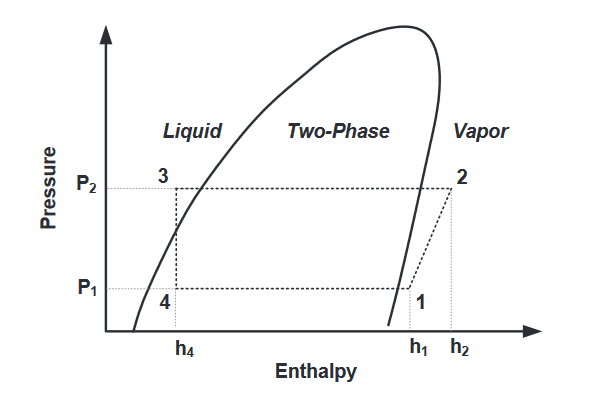}
\caption{Vapor Cycle on Pressure-Enthalpy diagram \cite{5991005}}
\label{fig_sim2}
\end{figure}

\textit{Background of the VCS:} The function of this system is to reduce and control the temperature of a medium by transferring its heat to the refrigerant (Evaporator) and then to transfer the heat from the refrigerant to an external medium (Condenser). The Compressor and the Expansion Valve bring the refrigerant fluid in the appropriate conditions of pressure and temperature for heat transfer. A typical refrigeration cycle comprises four stages through which the refrigerant undergoes: compression (from point 1 to point 2), condensation (from point 2 to point 3), expansion (from point 3 to point 4), and evaporation (from point 4 back to point 1) \cite{arora2000refrigeration,5991005}. This system is represented in Figure
 \ref{fig_sim1}.
 The cycle is defined by these transitions that are represented in Figure \ref{fig_sim2}. 
    VCSs are used in aircraft for cooling the cabin and the motors down. For control purposes, it is necessary to accurately measure the mass flow through the compressor with a small response time, lest it could lead to control instability \cite{zabczyk_mathematical_2020} and surge phenomenon \cite{Massiquet_2022}.
    This mass flow can be expressed as the ratio of the power supplied to the fluid by the enthalpy difference between the outlet and the inlet of the compressor \cite{reg2} :
    \[
  \dot{m} = \frac{C\omega(1-\alpha_\textit{loss})}{h_2 - h_1} = \frac{C\omega(1-\alpha_\textit{loss})}{h(T_\textit{out},P_\textit{out}) - h(T_\textit{in},P_\textit{in})}
 ~~(\star) \]

 \noindent where $h(P,T)$ is the enthalpy of the refrigerant fluid in vapor phase at temperature $T$ and pressure $P$. $\alpha_\textit{loss}$ is the proportion of dissipated energy and cannot be measured. 

\textit{On the ML-based virtual sensor:} 
 There are two standard physical mass flow sensors, the Coriolis \cite{baker_2016} and Venturi \cite{baker_20162} flowmeter. The first one is highly accurate but not robust to vibrations and hence not adapted to aircraft systems. The second one requires adding a bottleneck that induces a loss of charge and deteriorates the efficiency of the system. 
  This leads to the need for a virtual sensor based on other embeddable sensors. Unfortunately, there are no obvious physical equations that give the flow in function of these measured quantities with sufficient accuracy. 

\begin{figure}[h]
\centering
\includegraphics[width=3.6in]{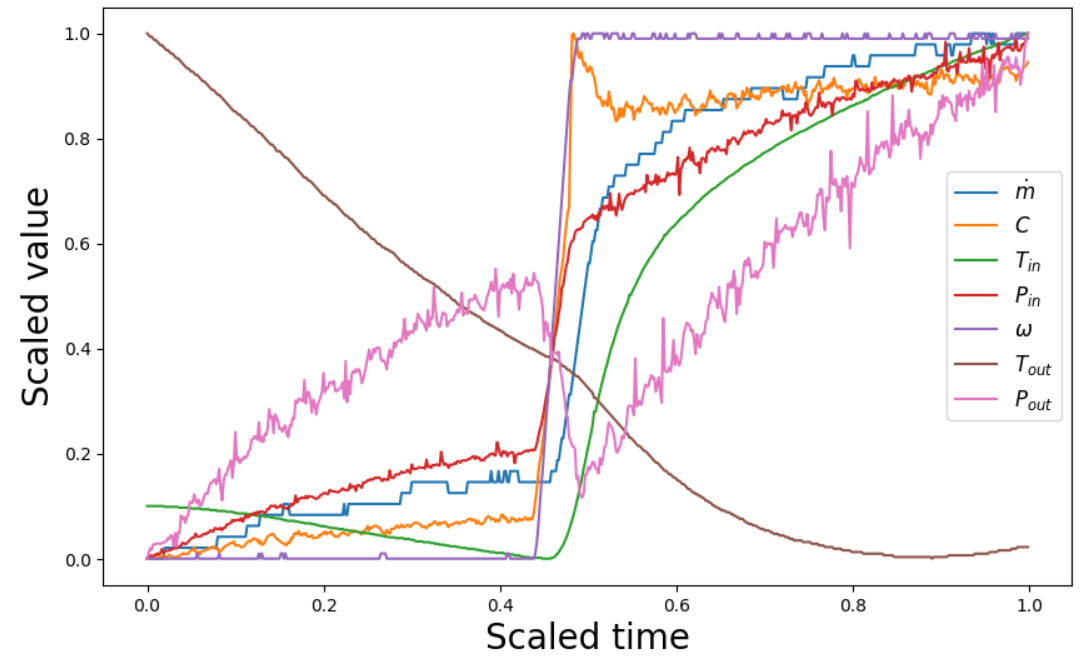}
\caption{Scaled measures}
\label{fig_sim3}
\end{figure}

 In the laboratory of LIEBHERR Aerospace in Toulouse, 
experiments were done on an industrial VCS. Several physicals
quantities were recorded at a high frequency during 30 hours,
including the mass flow measured with a Coriolis flowmeter. Indeed, even though the Coriolis flowmeter is not embeddable in aircraft, it is an excellent sensor for measuring the mass flow during the experiment in the laboratory.
Those measures pave the way to adopt a supervised machine learning framework 
to predict the mass flow from other physical
quantities (defined in the Nomenclature and exposed in Figure \ref{fig_sim3}).
Similar experiments were done in \cite{reg2} without the torque measure $C$ and in \cite{cnn} with additional data on the Condenser and the Expansion Valve. 

\textit{State-of-the-art and main contributions:}
In \cite{reg2}
an evaluation of statistical models based on Polynomial Regression (PR) is conducted. 
A CNN based on 2d convolutional networks (usually used to analyze images) 
is tested in \cite{cnn} and it is shown to outperform the PR models. 
The issue with the existing CNN approach is that no guarantee of 
physical properties are analyzed. 
In this article, we propose a different CNN architecture 
to guarantee time-translational invariance, 
insensitive to the order of input features, 
and causal properties for real-time estimation. 
The architecture is adapted from \cite{DBLP:journals/corr/abs-1901-10738}
for unsupervised learning of time series, with applications 
to classification and regression problems.




By expanding the scope of applications, 
one finds in the literature several models for designing virtual sensors \cite{8079133,dixon_fluid_1998,9724134,DBLP:journals/corr/abs-1901-10738,9127505}.
In \cite{9724134}, a CNN with recurrent skip connections is trained on simulated data, fine-tuned on real data to estimate physical quantities in electrical induction motors, and evaluated with electrical engineering metrics. 

The main contribution of this article is to show that
the proposed CNN significantly outperforms the PR model 
on the novel LIEBHERR dataset. 
To compare the difference between these models, 
we further propose a semi-automatic evaluation method 
to compute similar engineering metrics as in \cite{9724134}. 
The novelty is that we work on a real dataset so
it is needed to introduce a segmentation step 
of time series based on peak detection. 


The rest of the article is organized as follows. 
In Section II, PR and the proposed CNN models are detailed. In Section III, a two-fold evaluation method is explained, integrating machine learning metrics and engineering ones. The engineer's metrics are computed after a segmentation step, that enables to compute a multitude of metrics and to perform statistics analysis.
Finally, the results are presented and discussed in Section IV.

\section{ML Models}
 
We aim to develop a model 8capable of predicting real-time mass flow $\dot{m}$ based on specific measurements. Although equation $(\star)$ relates these variables, the quantity $\alpha_\textit{loss}$ is absent from our measurements. To circumvent this lack of physical relationship, we will rely on experimental data to construct two statistical models: Polynomial Regression (PR) and Convolutional Neural Networks (CNN). Each recording consists of a variable-length, multivariate time series containing seven measures, as represented in Figure \ref{fig_sim3}.
 \\

\textit{Polynomial Regression:} Statistical models are used in thermodynamics to determine the state function of a system at equilibrium from experiments. These models are called maps, and a standard way to create them is Polynomial Regression (PR) \cite{Rasmussen2000REVIEWOC}. Our baseline approach is PR as in \cite{cnn,reg2,Rasmussen2000REVIEWOC} and hence the model is static. It involves fitting a second-order polynomial function without cross-terms to the data points obtained from experimental recordings. This function was then used to predict the mass flow on the test recordings for evaluation. It is important to note that the prediction at time $t$ depends only on the measures at time $t$.
\[ \hat{\dot{m}}_{t} = \alpha+\sum_{i=1}^{6} \beta_{i} x^{(i)}_{t} + \sum_{i=1}^{6} \gamma_{i} \left(x^{(i)}_{t}\right)^{2},\]
where $\hat{\dot{m}}_t$ is the estimated mass flow at time $t$, \\
$x_t = (C_t,\omega_t,P_{\textit{in},t},T_{\textit{in},t},P_{\textit{out},t},T_{\textit{out},t})$ is the vector of input measures at time t, $\alpha\in\mathbb{R}$ and $\beta, \gamma\in\mathbb{R}^6$ are the model parameters.
  \newline 
  In \cite{reg2}, the authors propose to use the equation $(\star)$ by estimating the missing value $\alpha_\textit{loss}$ with a polynomial of other measures and found more precise and consistent results. Combining physical and statistical approach could be interesting, but this is not in the scope of this paper.
  


\noindent
\\

\textit{Convolutional Neural Network \cite{deep,lecun2015deep}:}
\noindent
A Convolutional Neural Network (CNN) is a series of convolution layers with adjustable coefficients, designed to minimize a loss function during the training process. 
We shall first specify the convolutional layer that we use in this paper and then analyze
its physical properties. We then specify the proposed CNN
architecture and discuss its advantage with respect to the standard
Polynomial Regression model.

Let $x\in \mathbb{R}^{I\times T}$ with $x^i_t$ the $i$-th feature at time $t\in\{1,...,T\}$. The causal one-dimensional convolution \textbf{(Conv1D)} is defined as  
\[
    y^{j}_t =\textit{Conv1D}(x)= \sum_{1\leq i\leq I} \sum_{s = 0}^{k-1} x^i_{t-s} w_s^{j,i},
\]
where $ y^{j}_t$ is the $j-$th output channel at time $t$, $I\in\mathbb{N}$ is the number of input channels, $k\in\mathbb{N}$ is the kernel size, $w^j\in\mathbb{R}^{I \times k}$ the kernels of the $j-$th output channel. 
This convolution is performed with zero padding, meaning that $x_t = 0$ for $t<0$. To compute the output at time $t$ the model uses the measurements $(x_{t-k+1},x_{t-k+2},...,x_{t})$ which are all anterior to $t$, this is a way to model causality \cite{oord2016wavenet}.
\\\noindent In \cite{cnn}, the CNN is based on two-dimensional convolution \textbf{(Conv2D)} defined as 
\[
    y^{j,l}_t =\textit{Conv2D}(x)= \sum_{i = 0}^{d-1} \sum_{s = 0}^{k-1} x^{l-\lfloor d/2 \rfloor +i}_{t-\lfloor k/2 \rfloor+s} w_s^{j,i},
\]
where $w^j\in\mathbb{R}^{k \times d}$ are the kernels of $j-$th output channel of size $k\times d$. There is one more exponent $'l'$ on $y$ because the convolution is also performed on the dimension of features.

\begin{figure}[h]
\centering
\includegraphics[width=1.82in]{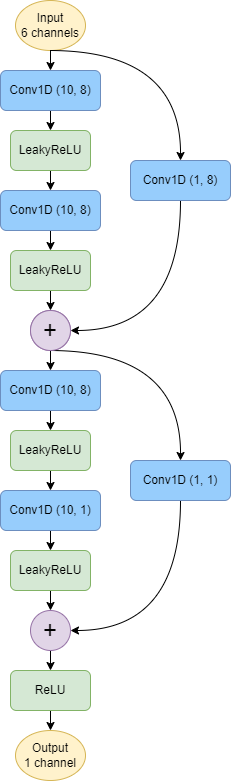}
\caption{\centering Feedforward CNN \newline \centering Conv1D (kernel size, number of output channels)}
\label{archi}
\end{figure}

In this paper, we propose a one-dimensional causal CNN with skip connections \cite{wu2020skip}. The complete architecture is detailed in Figure~\ref{archi} and is similar to the one used in \cite{DBLP:journals/corr/abs-1901-10738}.
A skip connection creates a shortcut from an early layer to a later one, linking the input of a convolutional block straight to its output \cite{wu2020skip}. Since various layers of a neural network capture distinct feature "levels," skip connections assist in preventing a drop in performance as more layers are added \cite{wu2020skip}. The CNN is trained with the MSE Loss. To accelerate the training, we also apply the weight normalization \cite{salimans2016weight} to re-parameterize the weights of each convolutional layer.




\noindent\textbf{Physical properties in CNN}
\begin{itemize}
\item Causality: Aiming to perform real-time estimation, the CNN has to be causal \cite{oord2016wavenet}. Thus, the model will only use present and past measures. 
\item Equivariance to time-translation: As demonstrated in \cite{lim2022equivariant}, CNNs are equivariant to translation in the dimension of convolution. Time-translation equivariance is an important property of physical systems, since the physical law does not evolve through time.

\item Insensitivity to input-feature order: The order of features in $x$ is arbitrary and has no physical meaning.

\noindent In causal Conv1D, the $j-$th output channel at time $t$ is
\[
    y^{j}_t = \sum_{1 \leq i \leq I} \sum_{s = 0}^{k-1} x^i_{t-s} w_s^{j,i}
\]
where $I$ is the number of input channels. 
If the input channel of $x$ is permuted (by $\pi$) to 
\[
    \tilde{x}^i_t = x^{\pi(i)}_t,
\]
then one can find an equivalent CNN with kernel 
$\tilde{w}^{j,i}_s = w_s^{j,\pi(i)}$
to have the same output $y^{j}_t$. 
This property does not hold in general if one uses the Conv2D layer. 
\end{itemize}
\noindent\textbf{Advantage of CNN over PR:} 
\begin{itemize}
\item Capturing Local Patterns\cite{oshea2015introduction}: CNNs are designed to capture local patterns and features within the data. In the context of time series, local patterns can represent short-term temporal dependencies, which are crucial for understanding the dynamics of the time series. PR, on the other hand, cannot effectively capture these local patterns as it does not use past measures for predictions.

\item Automatic Feature Extraction\cite{oshea2015introduction}: During the training step, CNNs automatically learn the relevant features from time series. PR uses directly the measures from sensors, which might not capture all the relevant information in the time series.

\item Handling Non-Linear Relationships\cite{oshea2015introduction}: Time series data often contains complex non-linear relationships between the input and output variables. CNNs with their multiple layers and non-linear activation functions can better model these non-linear relationships compared to the PR, which is inherently limited to polynomial relationships.
The activation functions of our model are LeakyReLU \cite{Maas2013RectifierNI} : \\
$$
\forall x\in \mathbb{R}, \textit{LeakyReLU}(x) = \left\{
    \begin{array}{ll}
        x & \mbox{if } x\geq 0\\
        0.1 x & \mbox{elsewhere}
    \end{array}
\right.
$$
\end{itemize}

PR has the advantage of being simple to train and easy to understand, as it only involves $13$ parameters. On the other hand, CNN are considerably more complex, requiring the tuning of $2,700$ parameters and involving more sophisticated computational processes.

\noindent 
\section{Evaluation}
\begin{figure}[h]
\centering
\includegraphics[width=3.6in]{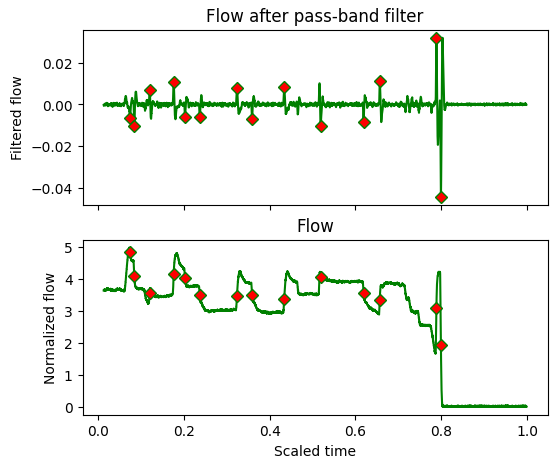}
\caption{First selection of candidates}
\label{first_step}
\end{figure}

\textit{EP Metrics:} To evaluate the performance of the ML methods, a dataset of systems measures and their corresponding mass flow was split into training (64\%), validation (18\%), and testing (18\%) sets. The training set was used to train the ML models, the model which had the smallest validation loss was selected, and the testing set was used to evaluate their performance.
The performance of the ML models was evaluated using the MSE and engineering performance (EP) metrics that are standard in engineering fields and well-suited for analyzing dynamical behavior \cite{9724134}. The EP metrics are defined as follows: \begin{itemize}
    \item $\Delta t_{80\%},\Delta t_{10\%}$\cite{9724134}: $80\%$ (resp.$10\%$) response time ($t_{80\%},t_{10\%}$) is the time value at which the
response signal has covered $80\%$ (resp.$10\%$) of the ramp amplitude. $\Delta t_{80\%}$ (resp. $\Delta t_{10\%}$)  is the absolute difference between the true and predicted $t_{80\%}$ (resp. $t_{10\%}$)

\item $\Delta t_{peak}$\cite{9724134}: Peak delay is the absolute delay between the predicted signal and the target one.
\item $\Delta{t_{\textit{conv}}}$: for a signal containing a ramp followed by a static state, $t_\textit{conv}$ is the time when the signal has converged to the final valued with 3\% of tolerance. $\Delta{t_{\textit{conv}}}$ is the delay between the predicted and true $t_{\textit{conv}}$.
\item $E_\textit{abs}$ (resp. $E_\textit{rel}$ ): Absolute (resp. relative) error on static states.
\end{itemize} 
In order to compute a lot of characteristic times and static errors (EP metrics), a first step of segmentation was developed to identify the ramps ($\Delta t_{80\%},\Delta t_{10\%},$ and $ \Delta t_\textit{conv}$), the overshoots and undershoots ($\Delta t_\textit{peak}$) and the static states ($E_\textit{rel}$ and $E_\textit{abs}$). Eventually, we have, for each of those metrics, several instances. To represent their distributions, we consider their 90\% quantile. 
\begin{figure}[h]
\centering
\includegraphics[width=3.6in]{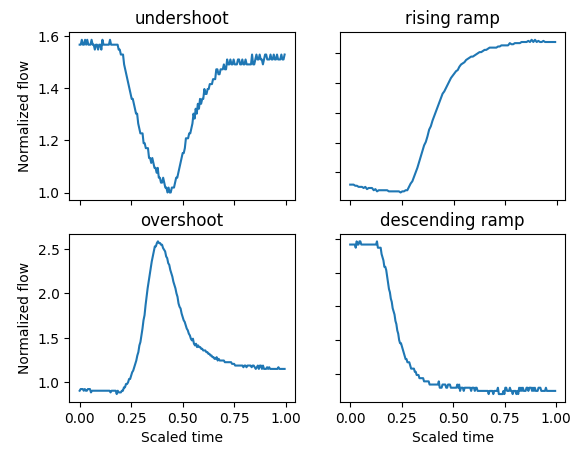}
\caption{Templates of patterns}
\label{patterns}
\end{figure}
\newline
\indent\textit{The segmentation process:} It consists of four steps. The patterns of interest are presented in Figure~\ref{patterns}. The first step consists in selecting windows in the dataset that will constitute a set of candidates for being one of the patterns. For this purpose, a Butterworth \cite{butter} pass band filter of order 6 is applied to the flow with system-specific cutting frequencies. The frequency peaks are detected via the scipy function \textit{'find\_peaks'} as shown in Figure~\ref{first_step}. The candidates are symmetric windows around the detected peaks.
 Then, a Dynamic Time Warping (DTW) k-means clustering \cite{articlekmeans} with 6 clusters is performed on the normalized signals (Figure~\ref{cluster} in Appendix). DTW, while not a true distance measure, serves as an effective metric for clustering patterns. Specifically, two signals exhibiting identical patterns, albeit with a mere time delay, will manifest a small DTW value. This is evident in Cluster 2 of Figure \ref{cluster}, where even if overshoots happen at varying times, they remain closely aligned in terms of DTW.
The identification of patterns is as follows:\begin{itemize}
    \item Cluster 1: Undershoots
    \item Cluster 2: Overshoots
    \item Cluster 3: Descending ramps
    \item Cluster 4: Rising ramps
\end{itemize}

 Cluster 5 and 6 correspond to other phenomenons like sinusoids. At this step, there are in clusters 1-4 some bad candidates. Aiming to remove those, we take the intersection with a DTW ball centered on the centers. A last manual selection is applied. Finally, we obtain satisfying numbers of each pattern (table II).

\begin{table}[h]
\centering
\begin{tabular}{cccc} \toprule

    \textbf{} &  train& test \\ \midrule
    Rising ramps&30&  8\\
    Descending ramps&35&   5\\
    Overshoots&79& 12\\
    Undershoots&22& 8\\
   \bottomrule

\end{tabular}

\caption{Number of patterns detected in each set}
\end{table}


\section{Results}
 Our study at LIEBHERR presents its results in a normalized manner due to the sensitivity and protection of the VCS data. Even so, the obtained results can provide a clear comparison between the performance of PR and CNN models. \\\indent As tabulated in Table III, the Convolutional Neural Network (CNN) model demonstrates superiority over the PR model both in terms of Machine Learning (ML) metric (MSE) and Engineering Performance (EP) metrics. The CNN model particularly excels in predicting static errors and $\Delta t_\textit{conv}$ (convergence time).\\\indent
As per the results, the CNN model outperforms the PR model by reducing the MSE from 0.18 to 0.06, the absolute error ($E_\textit{abs}$) from 0.73 to 0.28, and the relative error ($E_\textit{rel}$) from 28\% to 14\%. Furthermore, it provides a more accurate prediction of $\Delta t_\textit{conv}$ and $\Delta t_{80\%}$. PR has better results on $\Delta t_{10\%}$ and less good results on $\Delta t_{peak}$, however the differences are not significant. Thus, a better dynamic behavior is captured by the CNN model. Examples of predictions on patterns of interest can be found in Figure~\ref{example} in Appendix.
\begin{table}[h]
\centering
\begin{tabular}{ccccc} \toprule

    \textbf{} & \multicolumn{2}{c}{\textbf{PR}} & \multicolumn{2}{c}{\textbf{CNN}}\\
    
    {} & {train} & {test}& {train} & {test} \\ \midrule
    MSE  &0.2& 0.18&\textbf{0.063} & \textbf{0.06}  \\\midrule
    $E_\textit{abs}$&0.53  & 0.73 &\textbf{0.19} & \textbf{0.28}   \\
    $E_\textit{rel}$&21\%  & 28\% &\textbf{11\%}  & \textbf{14\%}  \\\midrule
    $\Delta t_{80\%}$&65  & 20 &\textbf{33}  & \textbf{17} \\ 
    $\Delta t_{10\%}$&92  & \textbf{3} &\textbf{17}  & 3.6 \\
    $\Delta t_\textit{conv}$ &97 & 44 & \textbf{74} & \textbf{14} \\
    $\Delta t_\textit{peak}$&\textbf{9}  & 7.1 & 11  & \textbf{7}    \\\bottomrule

\end{tabular}

\caption{Comparaison on MSE and 90\% quantile}
\end{table}



\section{Conclusion}

The study demonstrates that ML algorithms, specifically CNNs, can effectively function as virtual sensors in estimating physical quantities such as mass flow in aircraft compressor systems. Compared to PR, the causal one-dimensional CNN model, with the ability to capture dynamic behavior and non-linear relationships, outperforms in terms of MSE and EP metrics. The computation of multiple EP metrics on a large real dataset was performed via a semi-automatic segmentation method proposed in this article. The contribution is, on one hand, the design of a CNN-based model for estimating the flow in an industrial dataset from LIEBHERR Aerospace. On the other hand, the engineering evaluation is performed on several instances of ramps, overshoots, and undershoots.

Future work will focus on improving these models by using physical knowledge.
For the virtual sensor of mass flow, the author of \cite{reg2} proposed to use an incomplete equation $(\star)$. There exists multiple other methods to integrate physical knowledge into ML Models \cite{meng2022physics}.


\bibliographystyle{IEEEtran}
\bibliography{biblio}

\onecolumn

\section*{Appendix}

\begin{figure}[hbt]
\centering
\includegraphics[width=6.8in]{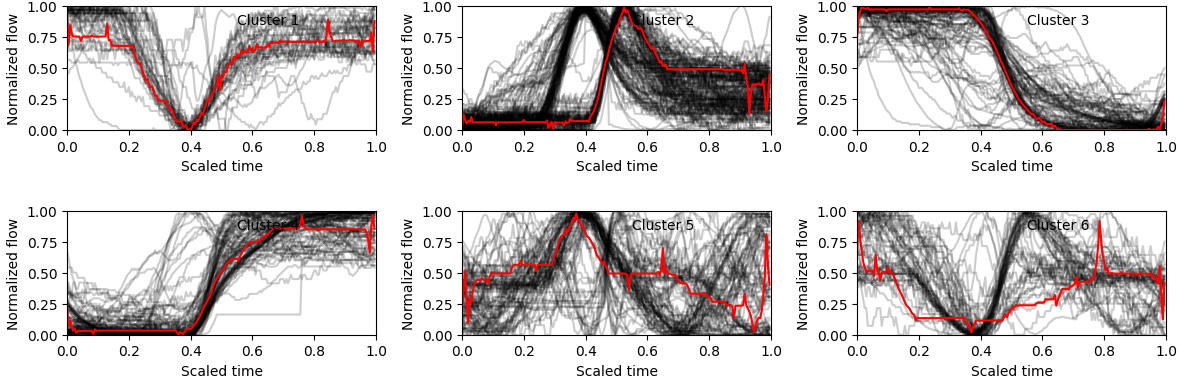}
\caption{Clusters with the DTW K-means algorithm. The centers are represented in red.}
\label{cluster}
\end{figure}

\begin{figure}[hbt]

\centerline{
\subfloat[$\Delta t_\textit{peaks}$]{\includegraphics[height=2.in]{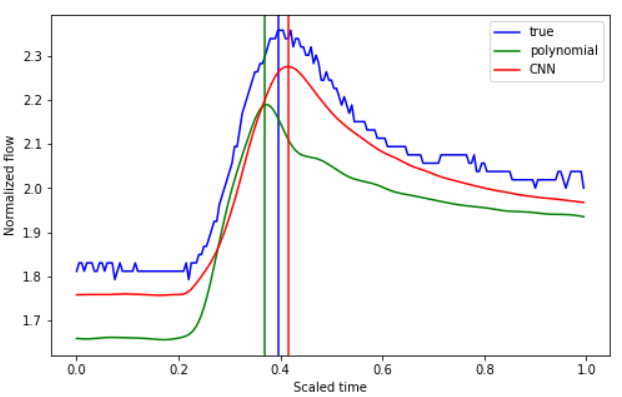}}
}
\subfloat[$\Delta t_\textit{conv}$]{\includegraphics[height=2.5in]{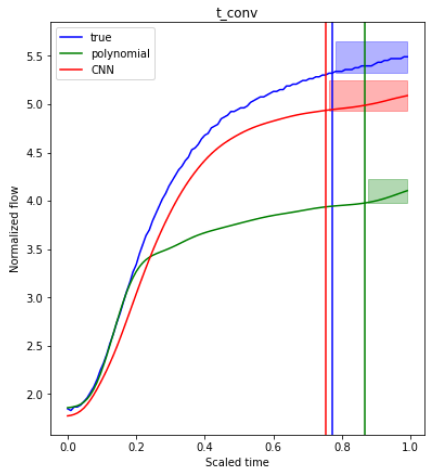}}
\hfil
\subfloat[$\Delta t_{10\%}$]{\includegraphics[height=2.5in]{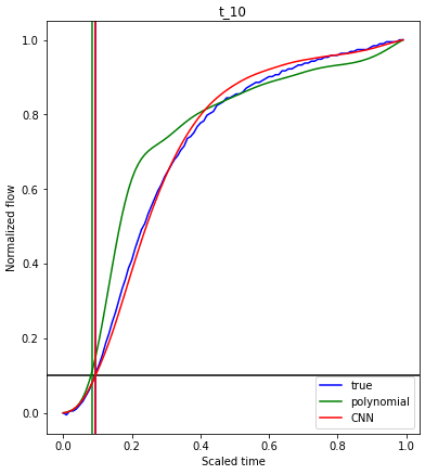}}
\hfil
\subfloat[$\Delta t_{80\%}$]{\includegraphics[height=2.5in]{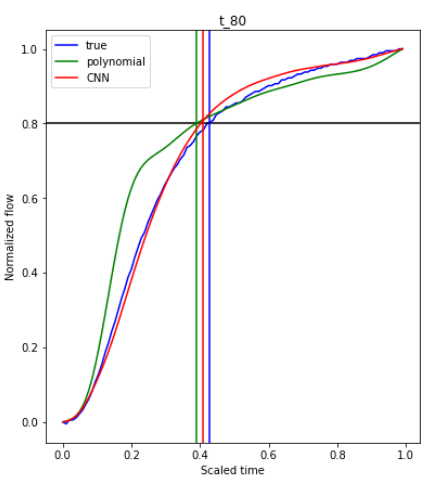}}\caption{On (a) $\Delta t_\textit{peaks}$ is computed from an overshoot. In (b) $\Delta t_\textit{conv}$ is computed from a rising ramp with the two predictions. On (c) and (d) there is the same ramp as (b) but the three signals are scaled in order to have the same initial and final value. In this manner it is easier to compute $\Delta t_{10\%}$ and $\Delta t_{80\%}$.}
\label{example}
\end{figure}

\end{document}